\documentclass[letterpaper, 10 pt, conference]{ieeeconf}
\usepackage{cite}
\usepackage{graphicx}
\usepackage[cmex10]{amsmath}
\usepackage{fixltx2e}
\usepackage{url}
\usepackage{widetext} 

\IEEEoverridecommandlockouts
\title{Bode's Sensitivity Integral Constraints: The Waterbed Effect in Discrete Time}
\author{%
Abbas Emami-Naeini$^{\star}$ and
Dick~de~Roover$^\star$
\thanks{$^\star$A. Emami-Naeini and Dick de Roover are with
SC Solutions, Inc., 1261 Oakmead Pkwy, Sunnyvale, CA 94085, USA.}%
\thanks{$^{\ast}$Corresponding author: Abbas  Emami-Naeini: emami@scsolutions.com.}
}
\begin{document}
\maketitle
\begin{abstract}
 Bode's sensitivity integral constraints define a fundamental rule about the limitations of feedback and is referred to as the waterbed effect. In a companion paper [35], we took a fresh look at this problem using a direct approach to derive our results. In this paper, we will address the same problem, but now in discrete time. Although similar to the continuous case, the discrete-time case poses its own peculiarities and subtleties. The main result is that the sensitivity integral constraint is crucially related to the locations of the unstable open-loop poles of the system. This makes much intuitive sense. Similar results are also derived for the complementary sensitivity function. In that case the integral constraint is related to the locations of the  transmission zeros outside the unit circle. Hence all performance limitations are inherently related to the open-loop poles and the transmission zeros outside the unit circle. A number of illustrative examples are presented.
\end{abstract}

\section{Introduction}
There is extensive literature on sensitivity of control systems and the fundamental and inevitable design limitations for linear time-invariant (LTI) systems [1]-[33]. One of the major contributions of Bode was to derive important fundamental and inescapable limitations on transfer functions that set limits on achievable design specifications. The majority of the previous results are based on Bode's sensitivity function,  $\mathcal{S}$, being the transfer function between the reference input to the tracking error or an output disturbance signal to the output (see Figure 1). Ideally we wish to have $|\mathcal{S}|\approx0$, which would provide perfect tracking and disturbance rejection. The sensitivity function is a measure of system sensitivity to plant variations [1]. In feedback control, the error in the overall transfer function gain is less sensitive to variations in the plant gain by a factor of $|\mathcal{S}|$ compared to errors in the open-loop control gain for those frequencies where $|\mathcal{S}|<1$. For a unity feedback system as in Figure 1 with the loop gain $L(z)$, $n$ poles and $m$ finite transmission zeros, the reference input $r$, the output $y$, and the tracking error $e$,
\begin{equation}
E(z)=(I+L(z))^{-1}R(z)=\mathcal{S}(z)R(z),
\end{equation}%

\begin{figure}
\centering
\includegraphics[width=3in]{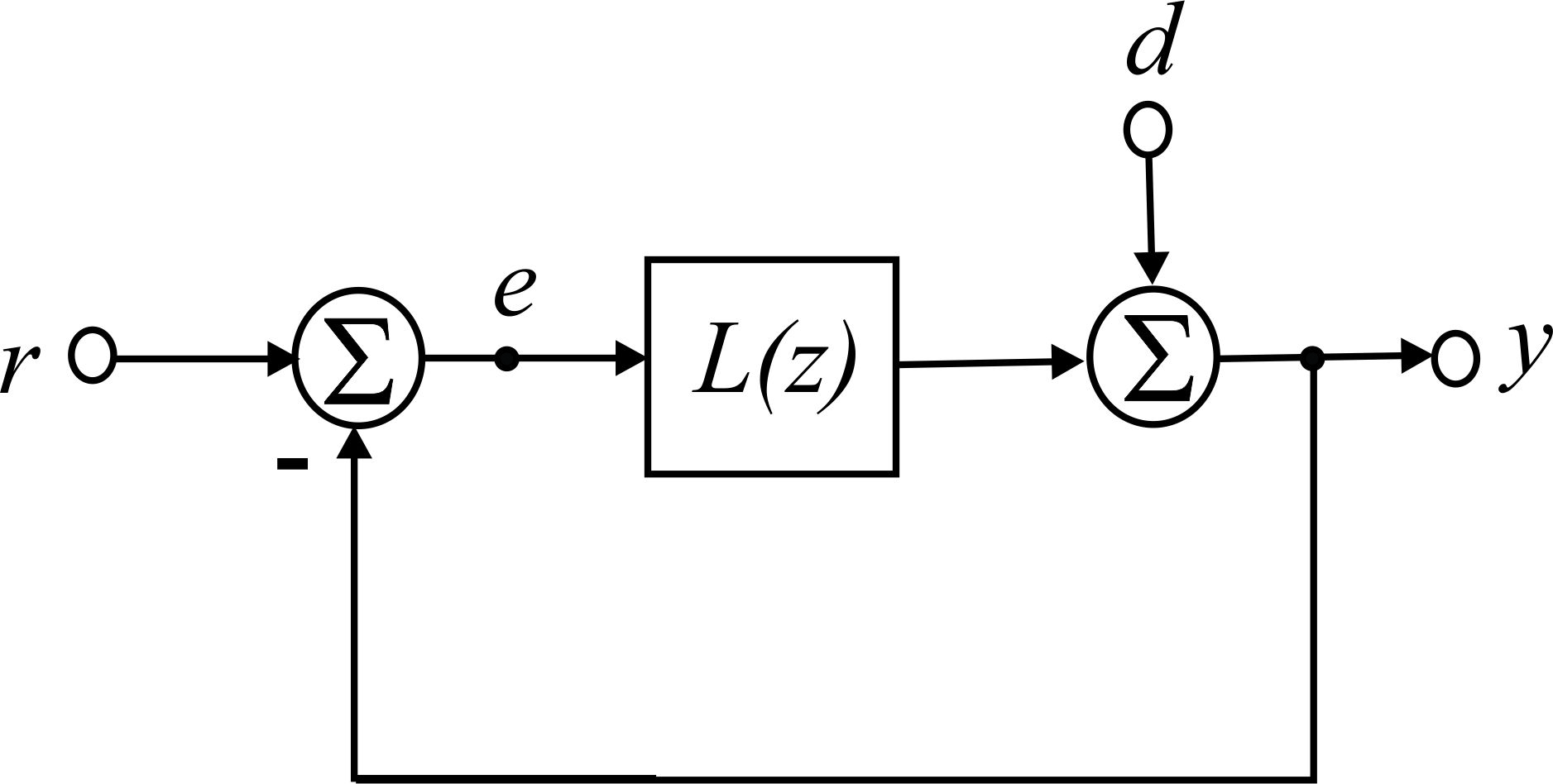}
\caption{Unity feedback discrete time system.}
\label{fig:fig_1}
\end{figure}
In addition to being the factor multiplying the system error, the sensitivity function, $\mathcal{S}$, is also the reciprocal of the distance of the Nyquist curve, $L(z)$, from the critical point (``-1" point). A large $|\mathcal{S}_{max}|$ corresponds to a Nyquist plot that comes close to the $-1$ critical point and a system having a small complex margin [1], [34] that comes close to the point of instability. The frequency based specification based on the above equation can be expressed as
\begin{equation}
	|E|=|\mathcal{S}||R|<e_{b}.
\end{equation}%
For minimum phase continuous time systems, the design rule was developed that the asymptotes of the Bode plot magnitude, which are restricted to be integral values for rational functions, should be made to cross over the zero-db line at a slope of $-1$ over a frequency range of about one decade around the crossover frequency [1]. In the discrete time case, the relationship between gain slope and phase does not hold [2]. However, it is approximately true for frequencies well below the Nyquist frequency. An alternative to the standard Bode plot as a design guide can be based on a plot of the sensitivity function as a function of frequency. In this format, we require that the magnitude of the sensitivity function, $|\mathcal{S}|$, be less than a specified value $|\mathcal{S}|<1/W_1$), over the frequency range $0\leq\omega\leq\omega_1$ for tracking and disturbance rejection performance, and that $|\mathcal{S}|\approx1$ over the range $\omega_2 \leq\omega$ for stability robustness. For proper open-loop stable discrete time systems [5],
\begin{equation}
\int_{0}^{2\pi}\ln(|\mathcal{S}(e^{j\omega})|)d\omega=0.
\end{equation}%
 Eq. (3) represents a fundamental trade-off relationship in feedback control. It implies that if we make the log of the sensitivity function very negative (where $|\mathcal{S}|<1$) over some frequency band to reduce errors in that band, then, of necessity, $\ln(|\mathcal{S}|)$ will be positive (where $ |\mathcal{S}|>1$) over another part of the band, and errors will be amplified there (see Figure 2). Note that this figure is a log-linear plot (not log-log Bode plot). This means that the effect of disturbances are reduced for frequencies where $|\mathcal{S}|<1$ and they are amplified (an undesirable situation) for frequencies where $|\mathcal{S}|>1$. This characteristic is referred to as the ``waterbed effect."  In Figure 2 we see that the area of disturbance attenuation is exactly balanced by the area of disturbance amplification as a result of Eq. (3).
\begin{figure}
\centering
\includegraphics[width=3in]{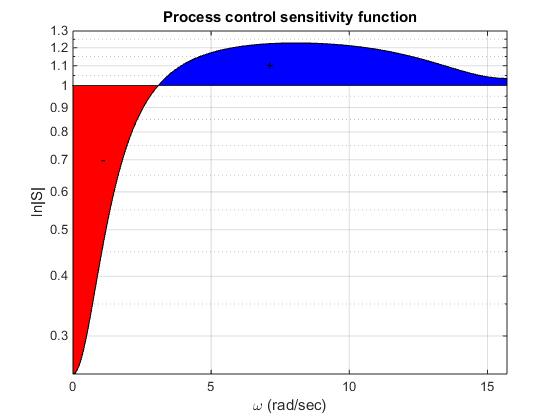}
\caption{Sensitivity function for process control example: positive and negative areas cancel exactly.}
\label{fig:fig_2}
\end{figure}
    In addition, there is a fundamental algebraic constraint given by,
\begin{equation}
	\mathcal{S}+\mathcal{T}=I,
\end{equation}%
where the complementary sensitivity function is defined as
\begin{equation}
	\mathcal{T}(z)=(I+L(z))^{-1}L(z),
\end{equation}%
which is the transfer function between the reference input $r$ and the output $y$ in Figure 1. Furthermore, at the poles and zeros of the loop gain $L(z)$ outside the unit circle, we must satisfy the interpolation conditions [33]
\begin{equation}
	|\mathcal{T}(z)|_{z=p_{i}}=1,\quad   |\mathcal{T}(z)|_{z=z_{i}}=0.
\end{equation}%
\begin{equation}
	|\mathcal{S}(z)|_{z=z_{i}}=1,\quad   |{\mathcal{S}}(z)|_{z=p_{i}}=0.
\end{equation}%
    Gunter Stein suggests that we think of Bode's integral constraints as a kind of a conservation law and for the lack of any better terminology he refers to it as ``conservation of sensitivity dirt," [31].  If performance improvements are sought in a frequency range, then deterioration of performance must be tolerated in another frequency range. In other words, ``there is no free lunch!" If there are unstable poles, the situation is worse, because the positive area where the sensitivity magnifies the error must exceed the negative area where the error is reduced by the feedback. If the system is minimum phase, then it is, in principle, possible to keep the magnitude of the sensitivity small by spreading the sensitivity increase over positive frequencies up to the Nyquist frequency, then the sensitivity function is constrained to take on a finite, possibly large, positive value at some point below the Nyquist frequency resulting in large peak in the sensitivity function.
\subsection{Extensions of Bode Sensitivity Constraints}
 Bode's results have been extended for the open-loop unstable case. The constraint shows that the integral of the sensitivity function is determined by the presence of poles outside the unit circle. Suppose the loop gain $L(z)$ has $n_{p}$ poles, $\{p_{i}\}$, outside the unit circle. References [5]-[8] showed that
\begin{equation}
	\int_{0}^{2\pi}\ln(|\mathcal{S}(e^{j\omega})|)d\omega=2\pi\sum\limits_{i=1}^{n_{p}}\ln(|p_{i}|).
\end{equation}%
The implications of Eq. (8) are the same as in continuous time except that the above integral is over a {\it finite} limit. If there are no poles outside the unit circle, then the integral is zero as before.
    If the system is not minimum-phase, the situation is even worse. An alternative to the above Eq. (8) is true if there is a nonminimum-phase zero of $L(z)$, a zero outside the unit circle. Suppose that the zero is located at $\beta_{0}=r_{0}e^{j\phi_{0}}, r_{0}>1$. Again, we assume there are $n_{p}$ poles outside the unit circle at locations $\alpha_{i}=r_{i}e^{j\phi_{i}}, r_{i}>1$ with conjugate values $\bar\alpha_{i}$. Now the condition can be expressed as a two-sided weighted integral [5]
\begin{equation}
	\int_{-\pi}^{\pi}\ln(|\mathcal{S}(e^{j\phi})|)W(r_0,\phi)d\phi=2\pi\sum\limits\limits_{i=1}^{n_{p}}\ln\left\vert {\frac{1-{\bar\alpha}_{i}\beta_{0}}{\beta_{0}-\alpha_{i}}}\right\vert,
\end{equation}%
where
\begin{equation*}
W(r_0,\phi)=\frac{r_{0}^2 -1}{r_{0}^{2}-2r_{0}\cos(\phi-\phi_{0})+1}.
\end{equation*}
The condition imposes a limitation on the sensitivity function due to the non-minimum phase zero. The constraint is especially severe if the non-minimum phase zero is near an unstable open-loop pole of the system (i.e. if $\beta_{i}\approx\alpha_{i}$). Based on this result, one expects especially great difficulty meeting both tracking and robustness specifications on the sensitivity with a system having poles and zeros close together outside the unit circle.
For the complementary sensitivity function [6] has shown that
\begin{equation}
\int\limits_{0}^{2\pi}\ln \left\vert \mathcal{T}(j\omega
)\right\vert d\omega =2\pi\left(
\sum\limits\limits_{i=1}^{i=n_{z}}\ln (\left\vert z_{i}\right\vert)+\ln(|K|)\right),
\end{equation}%
where $K$ is the first non-zero Markov parameter of $L(z)$. If $L(z)$ has the state space realization given by
\begin{eqnarray}
{\bf x}_{k+1}&=&{\bf Ax}_{k}+{\bf Bu}_{k},\\
{\bf y}_{k}&=&{\bf Cx}_{k},
\end{eqnarray}
then
\begin{equation}
K={\bf C}{\bf A}^{i-1}{\bf B},\quad i\geq 1,\quad K\neq 0.
\end{equation}
 These results and their extensions have been the subject of intensive study and have provided great insight into the problem.  Our derivation is direct and much simpler and does not rely on either Cauchy or Poisson-Jensen formulas that have been the focus of previous approaches to this problem.

    The organization of the rest of this Paper is as follows. In Section II we derive two fundamental relationships for the scalar case. One is a constraint on the sensitivity function and the other is a constraint on the complementary sensitivity function. Section III contains two SISO examples. The same results are derived for the multivariable systems in Section IV. Section V provides an illustrative MIMO example.  Concluding remarks are in Section VI. The proofs of the theorems are contained in the Appendices.
\section{Sensitivity Constraints for SISO Systems}
In this section we present two theorems that establish constraints on the sensitivity and complementary sensitivity functions for single-input single-output (SISO) discrete time systems and present several illustrative examples to show the merits of our results.\\
{\bf {Theorem 1}:} For any SISO closed-loop stable proper rational linear time-invariant (LTI) discrete system Bode's integral
constraint may be described as%
\begin{equation}
\int\limits_{0}^{2\pi}\ln \left\vert \mathcal{S}(e^{j\omega} )\right\vert
d\omega\mbox{}\\
=\left\{ 
\begin{array}{cl}
0 , &\text{OLS,} \\ 
2\pi\sum_{i=1}^{n_p}\ln\left\vert p_{i}\right\vert ,&  \text{OLU,}%
\end{array}%
\right. 
\end{equation}%
where $\mathcal{S}$ is the sensitivity function, $%
\{p_{i}\}$, $i=1,2,\ldots,n$ are the locations of the open-loop poles, and there
are possibly $n_{p}$ unstable open-loop poles at $\{p_{i}\}, i=1,...,n_{p}$
(including multiplicities) with $\{\vert p_{i}\vert \}>1$. OLS refers to an open-loop stable system and OLU refers to an open-loop unstable system.\\
\begin{proof}
See Appendix A. The proof is similar to the one in [10].
\end{proof}
The fundamental relationship is that the sum of the areas underneath the $
\ln(\left\vert \mathcal{S}\right\vert )$ curve is related to unstable open-loop poles of the system. If the system is open-loop stable, the areas cancel exactly. If the system is open-loop unstable then
additional positive area is added leading to further sensitivity
deterioration.\\
{\bf{Theorem 2}:} For any SISO closed-loop stable proper rational LTI discrete time system the complementary sensitivity integral constraint may be described by
\begin{equation}
\int\limits_{0}^{2\pi }\ln \left\vert \mathcal{T}(e^{j\omega}
)\right\vert d\omega =2\pi \left(\sum\limits_{i=1}^{i=n_{z}}\ln(\left\vert z_{i}\right\vert)+\ln(\left\vert K\right\vert)\right),
\end{equation}
and there are possibly $n_z$ non-minimum
phase transmission zeros of the system (including multiplicities) with $\left\vert\{z_{i}\}\right\vert>1$.\\
\begin{proof}
See Appendix B.
\end{proof}
The above results are consistent with those in [5]-[8] and [10].
\section{SISO Examples}
We present two SISO examples to illustrate the results.
{Example 1}:
Consider the process control system with the loop gain [2, page 257]%
\begin{equation*}
L(z)=\frac{0.2628(z-0.7)(z+0.8752)}{(z+0.5)(z-0.8187)^2}.
\end{equation*}%
The sensitivity function is,%
\begin{equation*}
{\mathcal{S}}(z)=(I+L(z))^{-1}=\frac{(z+0.5)(z-0.8187)^2 }{(z+0.4101)(z-0.6424\pm j0.1092)},
\end{equation*}%
\begin{equation*}
\left\vert \mathcal{S}(e^{j\omega} )\right\vert =\left\vert\frac{(e^{j\omega}+0.5)(e^{j\omega}-0.8187)^2}{(e^{j\omega}+0.4101)(e^{j2\omega}-1.2848e^{j\omega}+0.4246)}\right\vert.
\end{equation*}%
and by direct computation we find
\begin{equation*}
\int\limits_{0}^{2\pi}\ln \left\vert \mathcal{S}(e^{j\omega})\right\vert
d\omega =0.
\end{equation*}%
Since the system is open-loop stable, using our formula, Eq. (14), we find the same answer.
A plot of the log magnitude of the sensitivity function is shown in Figure 2. In this case the sensitivity function crosses unity
at $\omega=3.08$ rad/sec and hence there will be sensitivity reduction below that frequency and deterioration of sensitivity above that frequency up to the Nyquist frequency $\pi/T$.
\begin{figure}
\centering
\includegraphics[width=3in]{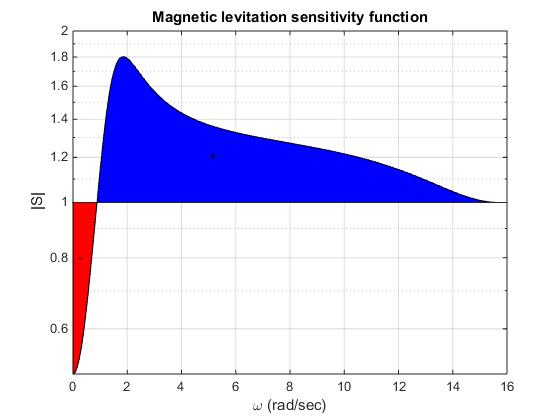}
\caption{Sensitivity function for Example 2.}
\label{fig:fig_3}
\end{figure}
The complementary sensitivity function is,
\begin{equation*}
{\mathcal{T}}(z)=1-\mathcal{S}(z)=\frac{0.2628(z-0.7)(z+0.8752)}{(z+0.4101)(z-0.6424\pm j0.1092)},
\end{equation*}%
\begin{equation*}
\left\vert \mathcal{T}(e^{j\omega})\right\vert =\left\vert\frac{0.2628(e^{j\omega}-0.7)(e^{j\omega}+0.8752)}{(e^{j\omega}+0.4101)(e^{j\omega}-0.6424\pm j0.1092)}\right\vert.
\end{equation*}%
By direct computation we have
\begin{equation*}
\int\limits_{0}^{2\pi}\ln \left\vert \mathcal{T}(e^{j\omega}
)\right\vert d\omega =-8.3966.
\end{equation*}%
Since the system is minimum phase, from Eq. (15), we find the same answer
\begin{equation*}
\int\limits_{0}^{2\pi}\ln \left\vert \mathcal{T}(e^{j\omega}
)\right\vert d\omega =2\pi\ln (|0.2628|)=-8.3966.
\end{equation*}%
{Example 2}:
Consider the magnetic levitation system with the loop gain [2, page 257]%
\begin{equation*}
L(z)=\frac{0.301(z-0.7)(z+1.0)}{(z+0.5)(z+0.8187)(z+1.2214)}.
\end{equation*}%
The sensitivity function is,%
\begin{equation*}
\mathcal{S}(z)=(I+L(z))^{-1}=\frac{(z+0.5)(z+0.8187)(z+1.2214)}{(z+0.3993)(z-0.8192\pm j0.2310)},
\end{equation*}%
\begin{equation*}
\left\vert \mathcal{S}(e^{j\omega})\right\vert =\left\vert\frac{(e^{j\omega}+0.5)(e^{j\omega}-0.8187)(e^{j\omega}-1.2214)}{(e^{j\omega}+0.3993)(e^{j\omega}-0.8192\pm j0.2310)}\right\vert.
\end{equation*}%
By direct computation we have
\begin{equation*}
\int\limits_{0}^{2\pi }\ln \left\vert \mathcal{S}(e^{j\omega})\right\vert
d\omega=1.2566.
\end{equation*}%
Since the system is open-loop unstable, using our formula, Eq. (14), we find the same answer:
\begin{equation*}
\int\limits_{0}^{2\pi }\ln \left\vert \mathcal{S}(e^{j\omega})\right\vert
d\omega =2\pi\ln(|1.2214|)=1.2566.
\end{equation*}%
A plot of the log magnitude of the sensitivity function is shown in Figure 3. The curve goes above unity at $\omega=0.865$ rad/sec. The sensitivity is reduced below that frequency and deteriorates from that frequency up to the Nyquist frequency $\pi/T$.
The complementary sensitivity function is
\begin{equation*}
{\mathcal{T}}(z)=1-\mathcal{S}(z)=\frac{0.301(z-0.7)(z+1.0)}{(z+0.3993)(z-0.8192\pm j0.2310)},
\end{equation*}%
\begin{equation*}
\left\vert \mathcal{T}(e^{j\omega})\right\vert =\left\vert\frac{0.301(e^{j\omega}-0.7)(e^{j\omega}+1.0)}{(e^{j\omega}+0.3993)(e^{j\omega}-0.8192\pm j0.2310)}\right\vert.
\end{equation*}%
\begin{equation*}
\int\limits_{0}^{2\pi}\ln \left\vert \mathcal{T}(e^{j\omega}
)\right\vert d\omega =-7.5439.
\end{equation*}%
Since the system is minimum phase, from Eq. (15),
\begin{equation*}
\int\limits_{0}^{2\pi }\ln \left\vert \mathcal{T}(j\omega
)\right\vert d\omega =2\pi\ln(|0.301|)=-7.5439.
\end{equation*}%
which is the same answer.\\
\section{Sensitivity Constraints for MIMO Systems}
In this section we present two theorems for multivariable discrete time systems.\\
{{\bf Theorem 3}:} For any square (and non-singular) LTI discrete time MIMO system with no hidden modes,
Bode's sensitivity integral constraint may be described as%
\begin{equation}
\int\limits_{0}^{2\pi }\ln \left\vert \det \left[ \mathcal{S}(e^{j\omega})%
\right] \right\vert d\omega =\left\{ 
\begin{array}{cc}
0, & \text{%
OLS} \\ 
2\pi\sum_{i=1}^{n_p}\ln\left\vert p_{i}\right\vert ,&  \text{OLU}%
\end{array}%
\right.
\end{equation}%
where $\mathcal{S}$ is the sensitivity function, $\{p_{i}\},  i=1,..., n$, are the locations
of the open-loop poles, and there are possibly $n_{p}$ unstable open-loop poles
(including multiplicities) with $\left\vert p_{i}\vert\right >1, i=1,\ldots,n_{p}$. OLS refers to an open-loop stable system and OLU refers to an open-loop unstable system.\\
\begin{proof}
See Appendix C.
\end{proof}
The authors believe that this is the first direct derivation of this result for the MIMO case. It is seen that the constraint on sensitivity, in this unweighted form,
is not dependent on the directions of the poles and their relative
interaction.\\
{{\bf Theorem 4}:} For any closed-loop stable square (and non-singular) LTI discrete MIMO system, with no hidden
modes, the integral constraint on the complementary sensitivity function may be
described as
\begin{equation}
\int\limits_{0}^{2\pi}\ln \left\vert \det \left[\mathcal{T}(e^{j\omega})\right] \right\vert d\omega = 2\pi \left(\sum\limits\limits_{i=1}^{i=n_{z}}\ln(|z_{i}|)+\ln(|K|)\right), 
\end{equation}
where $\{z_{i}\}, i=1,\ldots, m$ (including multiplicities) are
the (finite) closed-loop transmission zeros and $\{z_{i}\}, i=1\ldots,n_{z}$ are the non-minimum
phase transmission zeros of the system with $\left\vert z_{i} \right\vert >1$.\\
\begin{proof}
See Appendix D.
\end{proof}

The authors believe that this is the first direct derivation of this result for the MIMO case. Again this shows that the constraint on the complementary sensitivity
function is not dependent on the pole/zero directions either. These results are consistent with those in [5]-[8]. We now
demonstrate the merits of the above results with a multivariable
example.
\section{MIMO Example}
{Example 3}: Consider the system with the loop gain
\begin{equation*}
L(z)=\left[\begin{array}{cc}
\frac{0.1}{z-0.9}& \frac{0.2}{z-0.7} \\ 
\frac{0.1}{z-0.9} & \frac{0.1}{z-0.9}%
\end{array}%
\right] ,
\end{equation*}%
that has open-loop poles at $0.9$, $0.9$, $0.7$ and a finite transmission zero at $1.1$.
The sensitivity function is 
\begin{equation*}
\begin{split}
&\mathcal{S}(z) =(I+L(z))^{-1}\\
&=\frac{1}{\Delta(z)}\left[ 
\begin{array}{cc}
\Delta_1(z)& -0.2(z-0.9)^2 \\ 
-0.1(z-0.7)(z-0.9) & \Delta_1(z)\\
\end{array}%
\right] , \\
\end{split}
\end{equation*}
\begin{equation*}
\Delta_1(z)=(z-0.7)(z-0.9)(z-0.8)
\end{equation*}
\begin{equation*}
\Delta(z)=(z-0.5730)(z-0.8635\pm j0.0692)
\end{equation*}
\begin{eqnarray*}
&\det [\mathcal{S}(z)] =&\frac{(z-0.9)^2(z-0.7)}{(z-0.5730)(z-0.8635\pm j0.0692)},\\
\end{eqnarray*}%
\begin{equation*}
\left\vert \det [\mathcal{S}(e^{j\omega})\right]\vert =\left\vert \frac{(e^{j\omega}-0.9)^2(e^{j\omega}-0.7)}{(e^{j\omega}-0.5730)(e^{j\omega}-0.8635\pm j0.0692)}\right\vert\\.
\end{equation*}%
By direct computation we find
\begin{equation*}
\int\limits_{0}^{2\pi}\ln \left\vert \det \left[ \mathcal{S}(e^{j\omega})%
\right] \right\vert d\omega =0.
\end{equation*}%
Using our formula, Eq. (16), we find the same answer
\begin{equation*}
\int\limits_{0}^{2\pi}\ln \left\vert \det \left[ \mathcal{S}(e^{j\omega})%
\right] \right\vert d\omega =0.
\end{equation*}
A plot of the log magnitude of the determinant of the sensitivity function
is shown in Figure 4. Note that $\ln \left\vert \det [%
\mathcal{S}(j\omega )\right\vert =1$ at $\omega =3.59$ rad/sec. The complementary sensitivity function is%
\begin{equation*}
\begin{split}
\mathcal{T}(z)&=(I+L(z))^{-1}L(z)\\
&=\frac{1}{\Delta(z)}\left[ 
\begin{array}{cc}
 0.1(z^2-1.7z+0.74)&0.2(z-0.9)^2\\ 
0.1(z-0.7)(z-0.9)& 0.1(z^2-1.7z+0.74)%
\end{array}%
\right] .
\end{split}
\end{equation*}%
\begin{equation*}
\det \left[ \mathcal{T}(z)\right] =\frac{-0.01(z-1.1)}{(z^3-2.3z^2+1.74z-0.43)}.
\end{equation*}%
\begin{equation*}
\left\vert \det \left[ \mathcal{T}(e^{j\omega})\right] \right\vert =\left\vert\frac{-0.01(e^{j\omega}-1.1)}{(e^{3j\omega}-2.3e^{2j\omega}+1.74e^{j\omega}-0.43)}\right\vert.
\end{equation*}%
By direct computation we find
\begin{equation*}
\int\limits_{0}^{2\pi }\ln \left\vert \det \left[ \mathcal{T}%
(e^{j\omega})\right] \right\vert d\omega =-28.3.
\end{equation*}
This is a non-minimum phase system and from Eq. (17),
\begin{equation*}
\int\limits_{0}^{2\pi }\ln \left\vert \det \left[ \mathcal{T}%
(e^{j\omega})\right] \right\vert d\omega =2\pi(\ln(|1.1|)+\ln(|0.01|))=-28.3.
\end{equation*}
\begin{figure}
\centering
\includegraphics[width=3in]{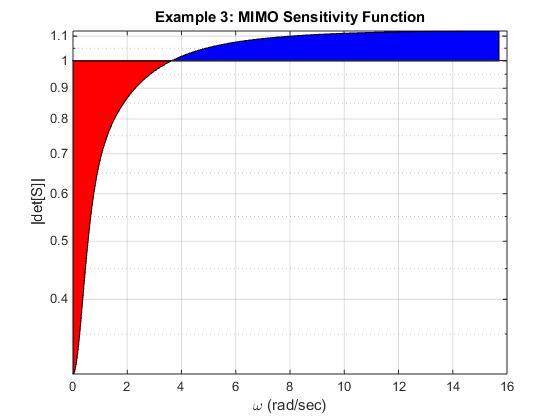}
\caption{Sensitivity function for Example 3.}
\label{fig:fig_7}
\end{figure}
\section{Conclusions}
In this paper, using a direct approach, we have taken a fresh look at the limitations imposed on
a discrete time feedback control due to constraints on the sensitivity function irrespective
of the feedback control synthesis techniques. We have shown that the
fundamental constraint on the sensitivity function is purely a function of unstable pole locations. The fundamental constraint
on the complementary sensitivity is a function of the transmission zeros outside the unit circle. The situation is made more difficult if there are 
poles/zeros outside the unit circle. \\
\textbf{Appendix A: Proof of Theorem 1.}\\
\begin{proof}
If the loop gain is denoted by $L(z)=\frac{N(z)}{D(z)}$, the sensitivity function is
\begin{equation}
\mathcal{S}(z)=\frac{1}{1+L(z)},
\end{equation}
and
\begin{equation}
\begin{split}
\int\limits_{0}^{2\pi }&\ln \left\vert \mathcal{S}(e^{j\omega} )\right\vert
d\omega =\int\limits_{0}^{2\pi }\ln \left\vert \frac{1}{1+L(e^{j\omega} )}%
\right\vert d\omega , \\
&= \int\limits_{0}^{2\pi }\ln \left\vert 1\right\vert d\omega
-\int\limits_{0}^{2\pi }\ln \left\vert 1+L(e^{j\omega} )\right\vert d\omega ,\\
&=-\int\limits_{0}^{2\pi }\ln \left\vert 1+\frac{N(e^{j\omega} )}{D(e^{j\omega} )%
}\right\vert d\omega ,   \\
&=-\int\limits_{0}^{2\pi }\ln \left\vert \frac{D({e^{j\omega}} )+N(e^{j\omega} )}{%
D(e^{j\omega} )}\right\vert d\omega ,   \\
&=\int\limits_{0}^{2\pi }\ln \frac{\left\vert D({e^{j\omega}} )\right\vert }{%
\left\vert D(e^{j\omega} )+N(e^{j\omega} )\right\vert }d\omega ,   \\
&=\int\limits_{0}^{2\pi }\ln \frac{\prod\limits _{i=1}^{i=n}\left\vert (e^{j\omega}
-p_{i})\right\vert }{\prod\limits _{i=1}^{i=n}\left\vert (e^{j\omega} -\tilde{p}%
_{i})\right\vert }d\omega ,   \\
&=\int\limits_{0}^{2\pi }\ln \frac{\prod\limits _{i=1}^{i=n}\left\vert (1-\left\vert p_{i}\right\vert e^{j(\phi_{i}-\omega)})\right\vert }{\prod\limits _{i=1}^{i=n}\left\vert (1-\left\vert \tilde{p}_{i}\right\vert e^ {j(\tilde\phi_{i}-\omega)})\right\vert }d\omega ,   \\
&=\int\limits_{0}^{2\pi }\ln \frac{\prod\limits _{i=1}^{i=n} (1+{\left\vert p_{i}\right\vert}^2-2\left\vert p_{i}\right\vert \cos {(\phi_{i}-\omega)})^{1/2}}{\prod\limits _{i=1}^{i=n} (1+{\left\vert \tilde{p}_{i}\right\vert}^2-2\left\vert {\tilde{p}_{i}}\right\vert \cos{(\tilde\phi_{i}-\omega)}%
)^{1/2}}d\omega ,   \\
&= \frac{1}{2}\int\limits_{0}^{2\pi }\sum\limits _{i=1}^{i=n}\ln (1+{\left\vert p_{i}\right\vert}^2-2\left\vert p_{i}\right\vert \cos {(\phi_{i}-\omega)})                                                           d\omega \\
& -\frac{1}{2}%
\int\limits_{0}^{2\pi }\sum\limits _{i=1}^{i=n}\ln (1+{\left\vert {\tilde p}_{i}\right\vert}^2-2\left\vert {\tilde p}_{i}\right\vert \cos {(\tilde\phi_{i}-\omega)}                           )d\omega .  
\end{split}
\end{equation}
We observe that the integrands above are periodic functions with a period $\pi$. Making a change of variables
\begin{equation}
\omega^{\prime}\buildrel\triangle\over =\omega-\phi_{i},\quad
\omega^{\prime\prime}\buildrel\triangle\over =\omega-\tilde\phi_{i}\\
\end{equation}
\begin{equation}
\begin{split}
&\int\limits_{0}^{2\pi }\ln \left\vert \mathcal{S}(e^{j\omega} )\right\vert
d\omega\\
&=\frac{1}{2}\sum\limits _{i=1}^{i=n}\int\limits_{-\phi_{i}}^{2\pi-\phi_{i} }\ln (   1+{\left\vert p_{i}\right\vert}^2-2\left\vert p_{i}\right\vert \cos {(\omega^{\prime}})                                                           )d\omega^{\prime} \\
& -\frac{1}{2}\sum\limits _{i=1}^{i=n}%
\int\limits_{-\tilde{\phi}_{i}}^{2\pi-{\tilde\phi}_{i} }\ln (1+{\left\vert {\tilde p}_{i}\right\vert}^2-2\left\vert {\tilde p}_{i}\right\vert \cos {(\omega^{\prime\prime})}                           )d\omega^{\prime\prime},\\
&=\frac{1}{2}\sum\limits _{i=1}^{i=n}\int\limits_{0}^{2\pi }\ln (   1+{\left\vert p_{i}\right\vert}^2-2\left\vert p_{i}\right\vert \cos {(\omega^{\prime}})                                                           )d\omega^{\prime} \\
&-\frac{1}{2}\sum\limits _{i=1}^{i=n}%
\int\limits_{0}^{2\pi }\ln (1+{\left\vert {\tilde p}_{i}\right\vert}^2-2\left\vert {\tilde p}_{i}\right\vert \cos {(\omega^{\prime\prime})}                           )d\omega^{\prime\prime}.  
\end{split}
\end{equation}
Using the integral identity%
\begin{equation}
\int_{0}^{2\pi}  \ln (1-2a\cos(x)+a^2)dx
 =\left\{
\begin{array}{cl}
0, \qquad\qquad a^2\le 1,\\
2\pi\ln(a^2),\quad a^2>1
\end{array}
\right.
\end{equation}%
we have
\begin{equation}
 \int\limits_{0}^{2\pi} \ln \left\vert \mathcal{S}(e^{j\omega} )\right\vert
d\omega =2\pi\sum_{i=1}^{n_p}\ln\left\vert p_{i}\right\vert.
\end{equation}%
Note that all the terms involving the (stable) closed-loop poles are zeros and the onlythe  non-zero terms are due to the unstable open-loop poles (those outside the unit circle). Hence the actual quantity is simply related to the sum of the magnitudes of the open-loop poles  outside the unit circle.
\end{proof}
\textbf{Appendix B: Proof of Theorem 2.}\\
\begin{proof}
The complementary sensitivity
function may be written as%
\begin{equation}
\mathcal{T}(z)=K\frac{(z-z_{1})(z-z_{2})...(z-z_{m})}{(z-\tilde{p}_{1})(z-%
\tilde{p}_{2})...(z-\tilde{p}_{n})},\qquad m\leq n
\end{equation}%
\begin{equation}
\begin{split}
&\int\limits_{0}^{2\pi } \ln \left\vert \mathcal{T}(e^{j\omega}
)\right\vert d\omega = \int\limits_{0}^{2\pi}\ln\frac{
\left\vert K\right\vert \prod\limits _{i=1}^{i=m} \left\vert(e^{j\omega}-z_{i})\right\vert }{\prod\limits _{i=1}^{i=n}\left\vert (e^{j\omega}-\tilde {p}_{i})\right\vert }d\omega , \\
&=\int\limits_{0}^{2\pi }    \ln \left\vert K\right\vert d\omega\\
&+\int\limits_{0}^{2\pi}           \sum\limits _{i=1}^{i=m}\ln (1+\left\vert z_{i}\right\vert^{2}-2\left\vert z_{i}\right\vert \cos(\omega-\phi_{i} ))^{1/2} d\omega ,  \\
&-\int\limits_{0}^{2\pi } \sum\limits _{i=1}^{i=n}\ln (1+\left\vert \tilde p_{i}\right\vert^{2}-2\left\vert \tilde p_{i}\right\vert \cos(\omega-\tilde\phi_{i} ))^{1/2}d\omega ,  \\
&=\int\limits_{0}^{2\pi }\ln \left\vert K\right\vert d\omega\\
&+\frac{1}{2}\int\limits_{0}^{2\pi}           \sum\limits _{i=1}^{i=m}\ln (1+\left\vert z_{i}\right\vert^{2}-2\left\vert z_{i}\right\vert \cos(\omega)) d\omega ,  \\
&-\frac{1}{2}\int\limits_{0}^{2\pi } \sum\limits _{i=1}^{i=n}\ln (1+\left\vert \tilde p_{i}\right\vert^{2}-2\left\vert \tilde p_{i}\right\vert \cos(\omega)) d\omega ,  \\
\end{split}
\end{equation}%
Using the integral identity%
\begin{equation}
\int_{0}^{2\pi} \ln (1-2a\cos(x)+a^2)dx
 =\left\{
\begin{array}{cl}
0, \qquad\qquad a^{2}\le 1,\\
2\pi\ln(a^{2}),\quad a^{2}>1.
\end{array}
\right.
\end{equation}%
we see that all the terms due to the stable closed-loop poles (inside the unit circle) and the transmission zeros inside the unit circle are all zeros. The only non-zero terms are due to the zeros outside the unit circle.
We then obtain
\begin{equation}
\int\limits_{0}^{2\pi }\ln \left\vert \mathcal{T}(e^{j\omega}
)\right\vert d\omega =2\pi \left(\sum\limits_{i=1}^{i=n_{z}}\ln(\left\vert z_{i}\right\vert)+\ln(\left\vert K\right\vert)\right).
\end{equation}
\end{proof}
\textbf{Appendix C: Proof of Theorem 3.}\\
\begin{proof}
\begin{equation}
\begin{split}
&\int\limits_{0}^{2\pi }\ln \left\vert \det \left[ \mathcal{S}(e^{j\omega})%
\right] \right\vert d\omega =\int\limits_{0}^{2\pi }\ln \left\vert \det %
\left[ I+L(e^{j\omega})\right] ^{-1}\right\vert d\omega\\
&\int\limits_{0}^{2\pi}\ln \left\vert \frac{1}{\det \left[ I+L(e^{j\omega})%
\right] }\right\vert d\omega
=-\int\limits_{0}^{2\pi}\ln \left\vert \det \left[ I+L(e^{j\omega})\right]
\right\vert d\omega . 
\end{split}
\end{equation}%
Suppose the loop gain $L(s)$ is written as an irreducible right-matrix
fraction description (MFD) [35, page 471] 
\begin{equation}
L(z)=N(z)D(z)^{-1},
\end{equation}%
\begin{equation}
\begin{split}
&\int\limits_{0}^{2\pi}\ln \left\vert \det \left[ \mathcal{S}(e^{j\omega})%
\right] \right\vert d\omega \\
&=-\int\limits_{0}^{2\pi}\ln \left\vert
\det \left[ I+N(e^{j\omega})D(e^{j\omega})^{-1}\right] \right\vert d\omega , \\
&=-\int\limits_{0}^{2\pi }\ln \left\vert \det \left[ D(e^{j\omega}
)+N(e^{j\omega})\right] \right\vert \det \left[ D(e^{j\omega})^{-1}\right] d\omega
,  \\
&=-\int\limits_{0}^{2\pi }\ln \left\vert \frac{\det \left[ D(e^{j\omega}
)+N(e^{j\omega})\right] }{\det \left[ D(e^{j\omega})\right] }\right\vert d\omega ,
 \\
&=-\int\limits_{0}^{2\pi}\ln \left\vert \frac{\det \left[ D(e^{j\omega}
)+N(e^{j\omega})\right] }{\det \left[ D(e^{j\omega})\right] }\right\vert d\omega ,
 \\
&=-\int\limits_{0}^{2\pi}\ln \left\vert \frac{\phi _{cl}(e^{j\omega})}{%
\phi _{ol}((e^{j\omega})}\right\vert d\omega ,  
\end{split}
\end{equation}%
where $\phi _{ol}(z)$ and $\phi _{cl}(z)$\ are the open-loop and closed-loop
characteristic polynomials of the system.%
\begin{equation}
\begin{split}
&\int\limits_{0}^{2\pi}\ln \left\vert \det \left[ \mathcal{S}(e^{j\omega})%
\right] \right\vert d\omega \\
&=\int\limits_{0}^{2\pi }\ln \frac{\prod\limits _{i=1}^{i=n}\left\vert (e^{j\omega}
-p_{i})\right\vert }{\prod\limits _{i=1}^{i=n}\left\vert (e^{j\omega} -\tilde{p}%
_{i})\right\vert }d\omega ,   \\
&=\int\limits_{0}^{2\pi }\ln \frac{\prod\limits _{i=1}^{i=n}\left\vert (1-\left\vert p_{i}\right\vert e^{j(\phi_{i}-\omega)})\right\vert }{\prod\limits _{i=1}^{i=n}\left\vert (1-\left\vert \tilde{p}_{i}\right\vert e^ {j(\tilde\phi_{i}-\omega)})\right\vert }d\omega ,   \\
&=\int\limits_{0}^{2\pi }\ln \frac{\prod\limits _{i=1}^{i=n} (1+{\left\vert p_{i}\right\vert}^2-2\left\vert p_{i}\right\vert \cos {(\phi_{i}-\omega)})^{1/2}}{\prod\limits _{i=1}^{i=n} (1+{\left\vert \tilde{p}_{i}\right\vert}^2-2\left\vert {\tilde{p}_{i}}\right\vert \cos{(\tilde\phi_{i}-\omega)}%
)^{1/2}}d\omega ,   \\
&= \frac{1}{2}\int\limits_{0}^{2\pi }\sum\limits _{i=1}^{i=n}\ln (1+{\left\vert p_{i}\right\vert}^2-2\left\vert p_{i}\right\vert \cos {(\phi_{i}-\omega)})                                                           d\omega \\
& -\frac{1}{2}%
\int\limits_{0}^{2\pi }\sum\limits _{i=1}^{i=n}\ln (1+{\left\vert {\tilde p}_{i}\right\vert}^2-2\left\vert {\tilde p}_{i}\right\vert \cos {(\tilde\phi_{i}-\omega)}                           )d\omega .  
\end{split}
\end{equation}
We observe that the integrands above are periodic functions with a period $\pi$. Making a change of variables
\begin{equation}
\omega^{\prime}\buildrel\triangle\over =\omega-\phi_{i},\quad
\omega^{\prime\prime}\buildrel\triangle\over=\omega-\tilde\phi_{i}\\
\end{equation}
\begin{equation}
\begin{split}
&\int\limits_{0}^{2\pi }\ln \left\vert \det \left[ \mathcal{S}(e^{j\omega})%
\right] \right\vert d\omega \\
&=\frac{1}{2}\sum\limits _{i=1}^{i=n}\int\limits_{-\phi_{i}}^{2\pi-\phi_{i} }\ln (   1+{\left\vert p_{i}\right\vert}^2-2\left\vert p_{i}\right\vert \cos {(\omega^{\prime}})                                                           )d\omega^{\prime} \\
& -\frac{1}{2}\sum\limits _{i=1}^{i=n}%
\int\limits_{-\tilde{\phi}_{i}}^{2\pi-{\tilde\phi}_{i} }\ln (1+{\left\vert {\tilde p}_{i}\right\vert}^2-2\left\vert {\tilde p}_{i}\right\vert \cos {(\omega^{\prime\prime})}                           )d\omega^{\prime\prime},\\
&=\frac{1}{2}\sum\limits _{i=1}^{i=n}\int\limits_{0}^{2\pi }\ln (   1+{\left\vert p_{i}\right\vert}^2-2\left\vert p_{i}\right\vert \cos {(\omega^{\prime}})                                                           )d\omega^{\prime} \\
&-\frac{1}{2}\sum\limits _{i=1}^{i=n}%
\int\limits_{0}^{2\pi }\ln (1+{\left\vert {\tilde p}_{i}\right\vert}^2-2\left\vert {\tilde p}_{i}\right\vert \cos {(\omega^{\prime\prime})}                           )d\omega^{\prime\prime}.  
\end{split}
\end{equation}
Using the integral identity Eq. (22)
we have 
\begin{equation}
 \int\limits_{0}^{2\pi} \ln \left\vert \det \left[ \mathcal{S}(e^{j\omega})%
\right] \right\vert d\omega =2\pi\sum_{i=1}^{n_p}\ln\left\vert p_{i}\right\vert.
\end{equation}%
Note that all the terms involving the (stable) closed-loop poles are zeros and the only non-zero terms are due to the unstable open-loop poles (those outside the unit circle). Hence the actual quantity is simply related to the sum of the magnitudes of the open-loop poles  outside the unit circle.
\end{proof}
\textbf{Appendix D: Proof of Theorem 4.}\\
\begin{proof}
Suppose the loop gain $L(z)$ is written as an irreducible right-matrix
fraction description (MFD) [35, page 471] 
\begin{equation}
L(z)=N(z)D(z)^{-1},
\end{equation}
then
\begin{equation}
\begin{split}
&\int\limits_{0}^{2\pi}\ln \left\vert \det \left[ 
\mathcal{T}(e^{j\omega})\right] \right\vert d\omega \\
&=\int\limits_{0}^{2\pi
}\ln \left\vert \frac{\det[N(e^{j\omega})]}{\det[D(e^{j\omega})+N(e^{j\omega })] }\right\vert d\omega , \\
&=\int\limits_{0}^{2\pi}\ln \left\vert \frac{\phi
_{z}(e^{j\omega})}{\phi _{cl}(e^{j\omega})}\right\vert d\omega ,  
\end{split}
\end{equation}
where the zeros of the $\phi _{z}(z)$ polynomial are the transmission zeros
of the system and $\phi _{cl}(z)$ is the closed-loop characteristic
polynomial.
\begin{equation}
\begin{split}
&\int\limits_{0}^{2\pi } \ln \left\vert \det\right[\mathcal{T}(e^{j\omega}
)\left]\right\vert d\omega = \int\limits_{0}^{2\pi}\ln\frac{
\left\vert K\right\vert \prod\limits _{i=1}^{i=m} \left\vert(e^{j\omega}-z_{i})\right\vert }{\prod\limits _{i=1}^{i=n}\left\vert (e^{j\omega}-\tilde {p}_{i})\right\vert} d\omega , \\
&=\int\limits_{0}^{2\pi }\ln \left\vert K\right\vert d\omega\\
&+\frac{1}{2}\int\limits_{0}^{2\pi}           \sum\limits _{i=1}^{i=m}\ln (1+\left\vert z_{i}\right\vert^{2}-2\left\vert z_{i}\right\vert \cos(\omega))d\omega ,  \\
&-\frac{1}{2}\int\limits_{0}^{2\pi } \sum\limits _{i=1}^{i=n}\ln (1+\left\vert \tilde p_{i}\right\vert^{2}-2\left\vert \tilde p_{i}\right\vert \cos(\omega)) d\omega ,  \\
\end{split}
\end{equation}%
Using the integral identity Eq. (26)
we see that all the terms due to the stable closed-loop poles (inside the unit circle) and the transmission zeros inside the unit circle are all zeros. The only non-zero terms are due to the transmission zeros outside the unit circle.
We then obtain
\begin{equation}
\int\limits_{0}^{2\pi }\ln \left\vert \mathcal{T}(e^{j\omega}
)\right\vert d\omega =2\pi \left(\sum\limits_{i=1}^{i=n_{z}}\ln(\left\vert z_{i}\right\vert)+\ln(\left\vert K\right\vert)\right).
\end{equation}
\end{proof}
\section*{Acknowledgments}
The authors gratefully acknowledge the help of Drs. S. A. McCabe and Jun-Kyu Lee of SC Solutions.

\end{document}